# VOLUME IMAGES VECTOR DISPLAY BASED ON A DIFFRACTIVE SCREEN

José J. Lunazzi[1] and M. Diamand [2]

[1] Physics Institute, [2] Faculty of Electrical Engineering, Campinas State University.

C.P. 6165, 13083-100 Campinas-SP-Brazil

A white light system based on a 65 cm x 35 cm diffractive screen is demonstrated to be capable of displaying three-dimensional figures with continuous horizontal parallax. Three computer-controlled mirrors and a diffractive-refractive optical system are employed for positioning each element of the figure. No visual accessories are necessary and more than one observer can watch it simultaneously.

**1. Introduction**

**2. Encoding of the z coordinate.**

**3. Decoding of the z coordinate by projection on the diffracting screen.**

**4. Depth of focus considerations.**

**5. Distorsion considerations.**

**6. Experimental details**

**7.Conclusions**

**References.**

**1. Introduction**

Attempts have been made to obtain by electronic means, images that can be viewed in 3D without glasses, having at least horizontal continuous parallax. Direct volume display devices have large fields of view but need massive moving components[1,2] and cannot generate straddling figures because they cannot be shown outside of the protective housing. Real-time computer-generated holography[3,4] requires enormous computing capability and generates small images. The authors[5),12)] have described the possibility of displaying under white light computer-generated 3D figures with continuous horizontal parallax, using the wavelength encoding of depth principle[6)].

It involves creating a horizontal line with a continuous white light spectrum proportionally wide to the coordinate to encode, which can be decoded rendering a replicated light point by means of a white-light holographic screen. Its practical demonstration, however, was limited to displaying a circle within a plane which appears oblique to the screen. This was because no angular-controlled mirrors were available at that time, for a purpose where galvanometers were intended. We performed one more step towards the practical demonstration of a completely controlled 3D vector figure by using three stepping motors to rotate mirrors which direct the light. The **x**,**y** position of each voxel (volume element) composing the figure is determined by deflecting a light beam through two controlled rotating mirrors.

A third controlled rotating mirror is necessary to determine the **z** coordinate in two steps: it is first encoded by means of a variable chromatic dispersion applied to a white-light beam, then a diffractive screen decodes it, placing the voxel in its three-dimensional position. It would be natural to think that the volume figures generated on this complex diffraction encoding and decoding system could be affected by some kind of distortion This does not happen, as we verified by generating very linear figures to be analyzed by stereophotography.

## 2. Encoding of the *z* Coordinate.

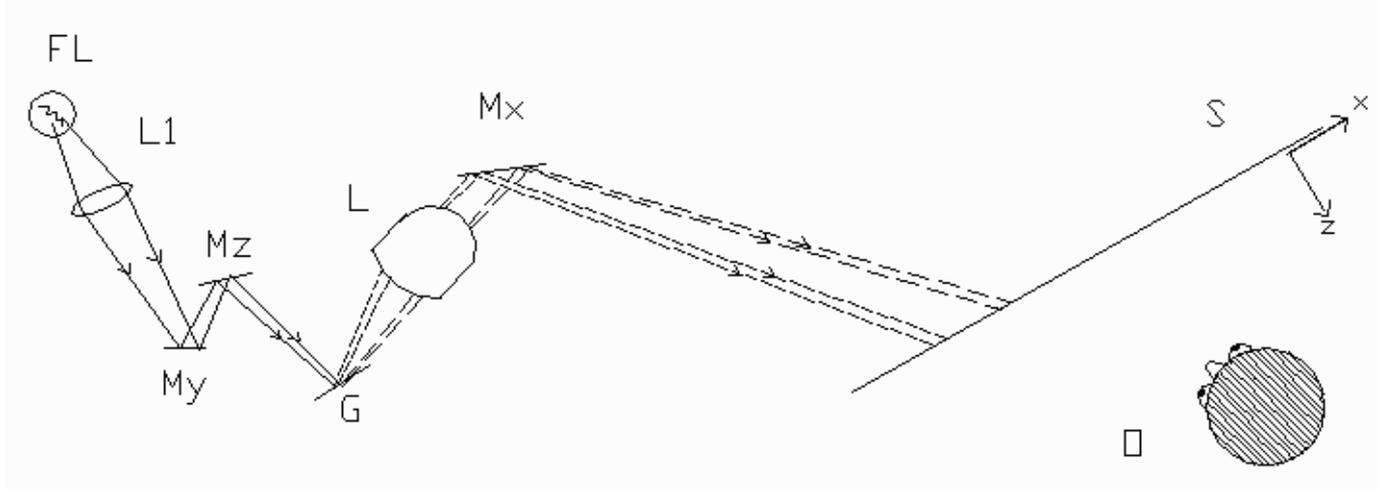

**Fig. 1** *Three-dimensional display based on diffraction encoding and decoding.*

The system is described in the scheme of Fig.1, showing a horizontal plane where a lens **L1** produces a small real image of a lamp filament **FL** close to a diffraction grating **G**, rapidly positioned horizontally by tilting mirror **Mz**, and vertically by tilting mirror **My**. Light diffracted from the grating is focused as a second multiwavelength image of the filament by lens **L** on the diffractive screen **S**, through reflection on mirror **M$_x$** whose rotation axis is vertical.

We consider as the image coordinates (**x,y**) the coordinates at the screen, **x** being the horizontal, **y** the vertical, while **z** is the coordinate normal to the screen. The movement of the filament's first image occurs within an approximately spherical surface that crosses the diffraction grating. This surface would be approximately a plane for mirrors **Mz** and **My, since it is** far from the grating. The center of this surface is located close to mirrors **Mz** and **My**, which are close to one another as in a conventional laser scanning system. The expression of the white light waves incident on the grating is:

$$w_i(x_g) = \exp(ikx_g \sin\theta) \exp\left[-\frac{ik(x_g^2 + y_g^2)}{2z_g}\right] \quad (1)$$

corresponding to waves of unitary amplitude and wave vector ☐ incident at the angle θ on the grating, where $x_g$ is the horizontal coordinate on the grating and $y_g$ the vertical coordinate. $z_g$ is the distance the image makes with the grating. The transmittance function $t(x_g)$ is:

$$t(x_g) = 1 + a \cos 2\pi v x_g \quad (2)$$

where ν is the spatial frequency of the grating. Employing a complex notation in which we also express equation (2) and selecting the terms corresponding to one diffracted order, the images obtained when the light is diffracted by the grating corresponds, at each specific wavelength, to:

$$w_d(x_g) = \frac{a}{2} \exp\left[2\pi x_g\left(\frac{\sin\theta}{\lambda} - v\right)\right] \exp\left[-\frac{ik(x_g^2 + y_g^2)}{2z_g}\right] \quad (3)$$

whose set of positions constitutes a spectral sequence[6] having distance $z_g$ from the illuminated region on the grating. This apparent blurring constitutes an encoding in the form of a spectral arc. As a result, we designed with our system a **z** coordinate encoder whose starting element consists of a thin white light beam locating its spot at different distances from a grating by a simple circular movement of a mirror. Figure 2 shows how it works with some colors diffracted in approximately a circular arc of radius $z_g$ [7,8].

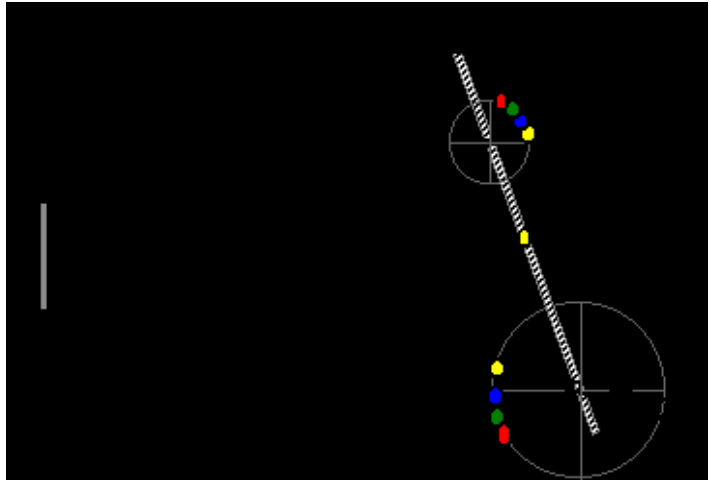

**Fig. 2** *Diffraction coding of the depth parameter.*

Three positions of the mirror are shown. The separate spot is the white light image of the lamp, the two assemblies of three spots indicates two spectral encoding arcs **SE** around the region of incidence on the grating **G**, one being made of real images while the other is composed of virtual images. For simplicity, this scheme represents a transmission grating, but the situation is entirely analogous for a reflection grating. Coordinate **z** is encoded by controlling the angular position of mirror $M_Z$. Its sign can be changed from positive to negative values very naturally and in continuous sequence when diffracted images of the source change from real to virtual because in the spectral sequence longer-to-shorter wavelengths reverse at the position corresponding to zero z value.

## 3. Decoding of the *z* coordinate by Projection on the Diffracting Screen.

The way in which the coded image is enlarged and decoded is shown in Figure 3 for three rays whose relative wavelengths are indicated by the size of the segments on its dashed lines. We did not considered in this figures the presence of the third mirror of the system, $M_x$, for simplicity.

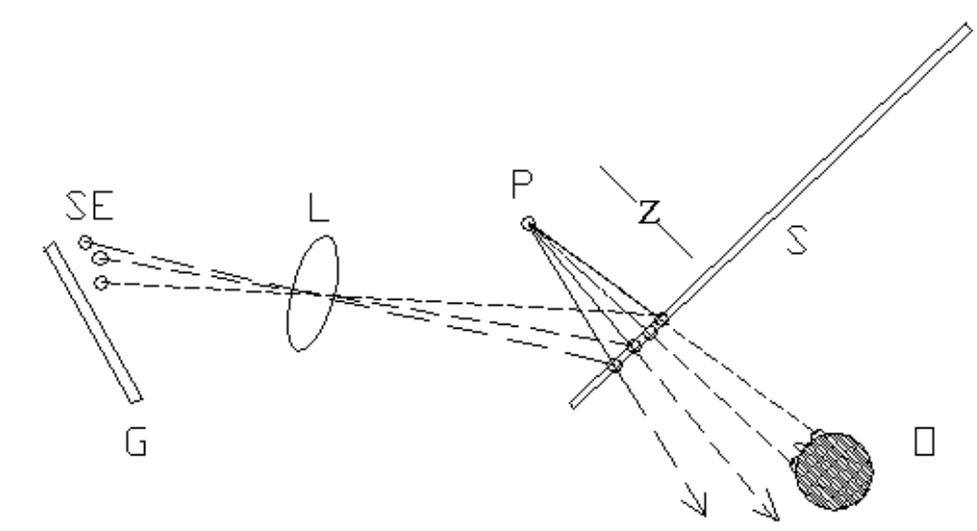

**Fig. 3(a)** *Generation of a virtual point.*

Figure 3a corresponds to the case where real images of the source are diffracted along the extension of **SE**. The spectral encoding is imaged on the diffractive screen **S** by lens **L**, creating a virtual image **P**. The properties of the holographic diffracting screen acting under white light were described in[9] considering it as the hologram of a point source, while in[10] the description included the extension of the vertical field of view, the screen being considered as the hologram of a vertical diffusing slit. We can understand that a certain coordinate value coming from the value $z_g$ is encoded as the lateral extension of **SE** and must be proportional to resulting coordinate **z** of the final image due to the geometrical proportionality of the situation. A larger extension of **SE** brings, after projection, a larger **z** value, then we see the proportionality of the whole encoding and decoding process.

The shortest wavelength ray was selected on the figure as the one that happens to reach the right eye of the observer, and another ray of different wavelength was added to the figure corresponding to the one reaching his left eye. It is clear that the visual system of the observer can perform a triangulation operation in the horizontal plane to represent the proper depth of the image. We must also consider that, although a ray tracing composition can easily show that in a vertical plane the screen acts as an ordinary diffusing screen and the image point comes from the screen and not from point **P**, the limited size of the eye's pupil and also the large size image point on the screen prevents the representation from appearing astigmatic at each voxel.

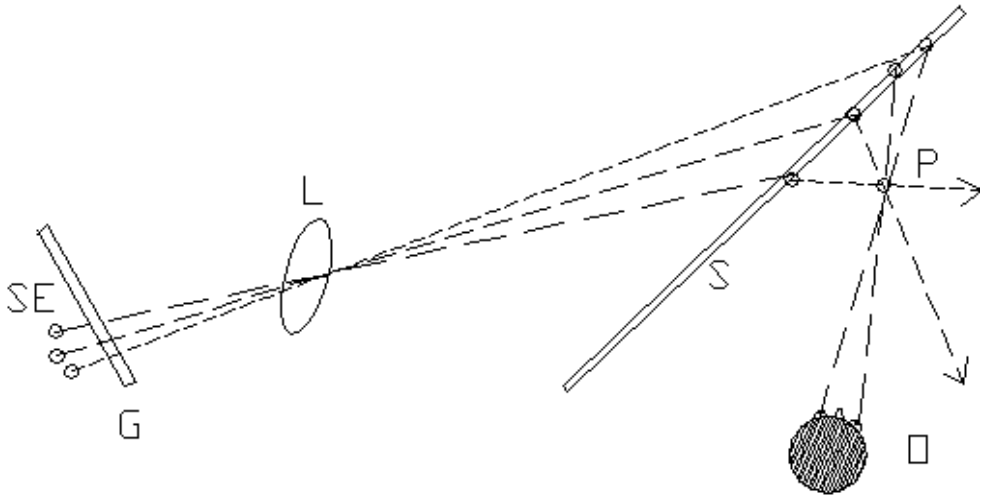

**Fig. 3(b)** *generation of a real point*

We see in Figure 3b the situation when the mirror $M_z$ is rotated to put the filament image before the grating, resulting in virtual diffracted encoded spectral images **SE** and a real final image for the observer. Figure 3 did not include mirror $M_x$, which is why **x** coordinate appears to be related to **z** coordinate. When a **z** coordinate value is indicated within the computer it acts to position mirror $M_z$ rendering the proper coding but giving to the **x** coordinate an initial value as a function of **z**, which is corrected by calculations acting on mirror $M_x$. So, the system renders to the image the corresponding **x** value. The observer can look around the image getting a complete analogous to the continuous horizontal parallax of the original object.

## 4. Depth of Focus Considerations

The whole encoding-decoding procedure implies a focusing condition to be fulfilled. It takes place between the surface giving the allowed positions of the diffracted images of the filament and the screen plane. We created a computer program to learn the numerical values of the coordinates of the image surfaces at the screen as a function of wavelength. We employed the values given below for the distances mirror-grating, grating-lens, lens-screen, and employed the classical gaussian formula for imaging by a lens from an object plane to an image plane. The diffraction equation of a grating was employed for this ray tracing, without any approximations.

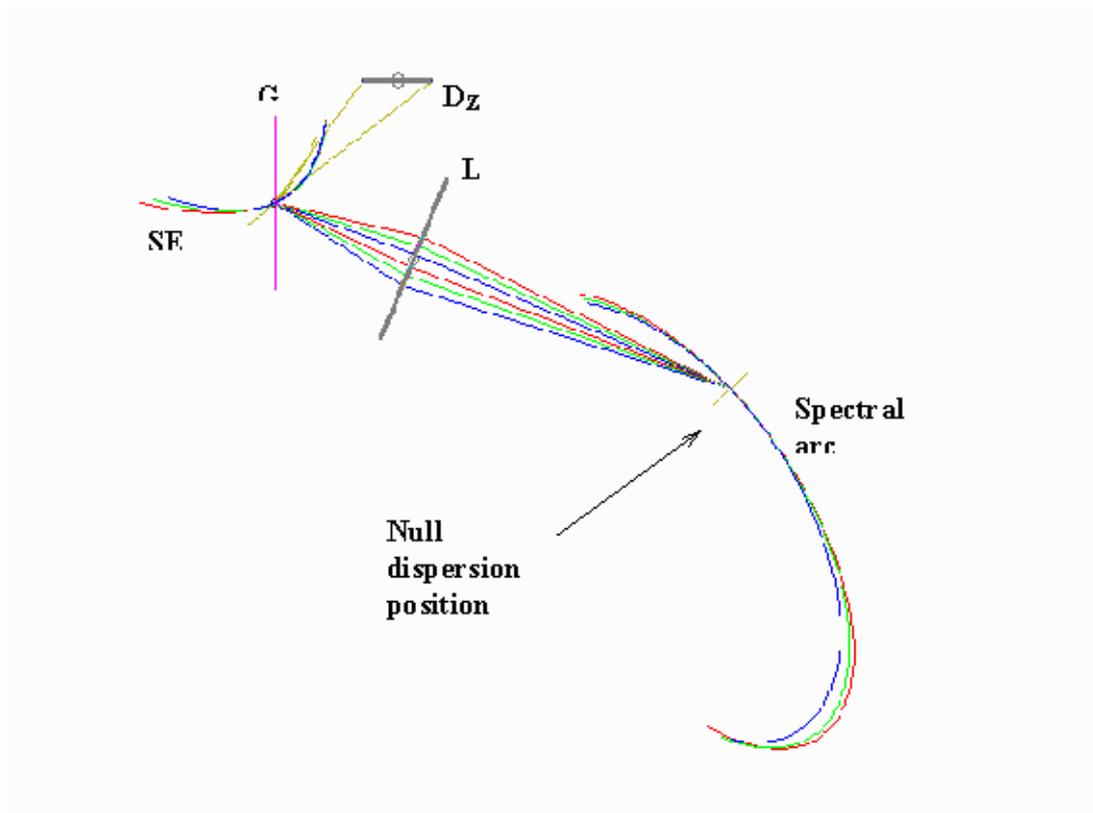

**Fig. 4** *Calculated distribution of wavelength positions after first diffraction and projection.*

The result is shown in Figure 4 where three wavelengths (red, green and blue) having two rays each are employed to indicate the converging action of the lens. When mirror $M_z$ is rotated, white reflected light generates the family of spectral arcs **SE** at the encoding step, and the final spectral arcs at the screen. These arcs are blue at the internal part, going towards the red wavelength which is at the external part. We can see that there is a portion of this family (we named it **"Spectral arc"**) which can be considered almost straight. The task of positioning and focusing the elements of this experiment, although difficult and delicate, can give a good result at the resolution values we employed.

When we add mirror $M_x$ to the situation of Figure 4 it appears that we lose the advantage of having a plane where focusing is accurate and can be closely matched to the plane of the screen. Fortunately, the depth of focus of our representation is large enough to cover all the displacements of the projected spectral arcs from one side of the screen to the other. As we describe below, this is because we are not dealing with resolution at its limit from the optical system, our minimal resolution is the size of the image of the filament on the screen, which is much larger. We show in Figure 5 the scheme in perspective of the complete system which includes the spreading of light from the screen to the observer.

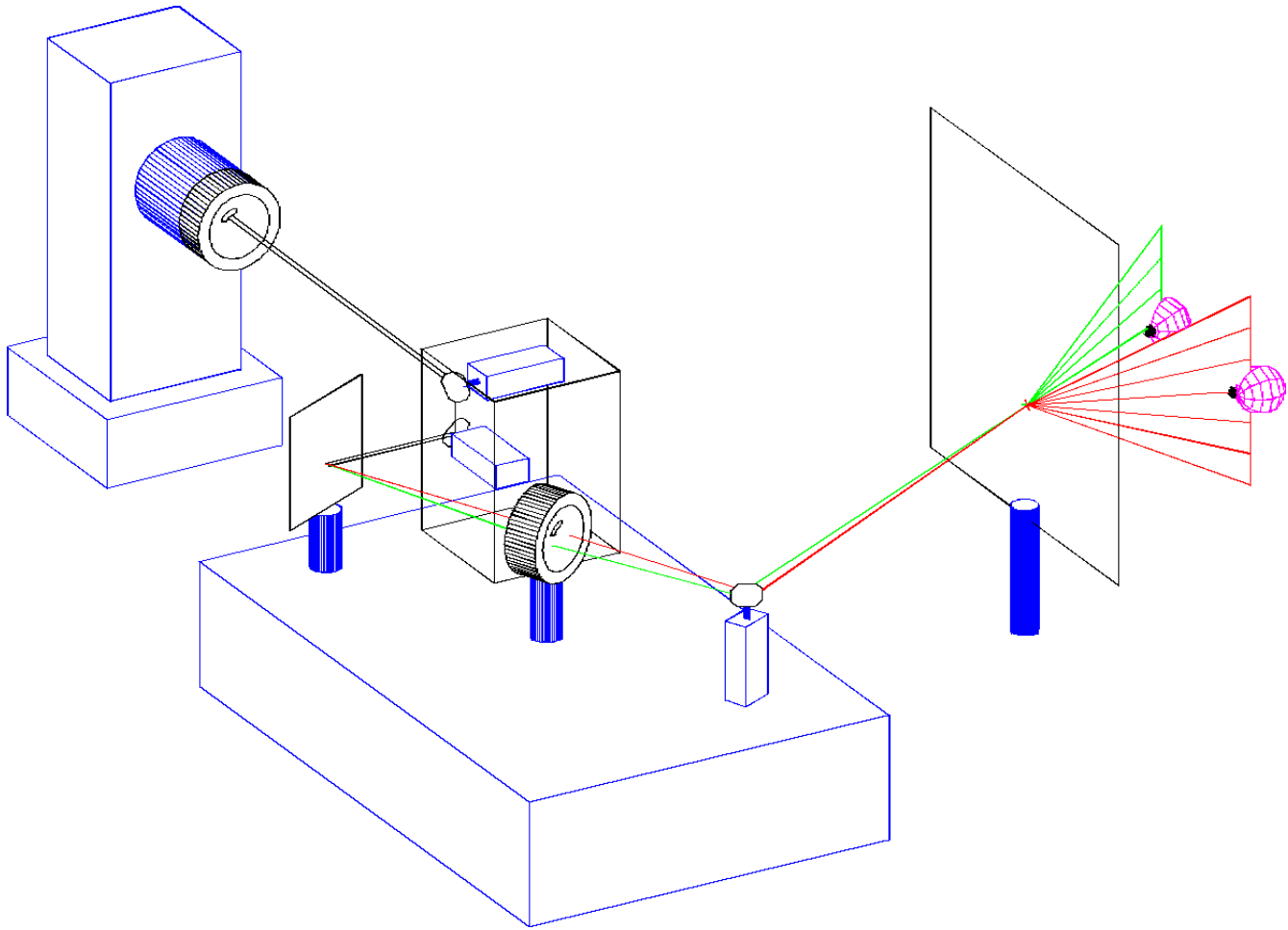

**Fig. 5** *Projection system. (a) The white light beam impinges at the center of the grating.*

In case a) the filament image is set precisely on the grating, and convergence of the final image is precisely on the screen. Considering that light impinges at a 45 degree angle to obtain the corresponding diffraction condition and that it is dispersed only in the vertical direction, it creates a continuous sequence of fans of light beams, a different wavelength and direction corresponding to each one. Two directions corresponding to two wavelengths are shown in this figure, where each eye will receive one fan coming from the same region of the screen.

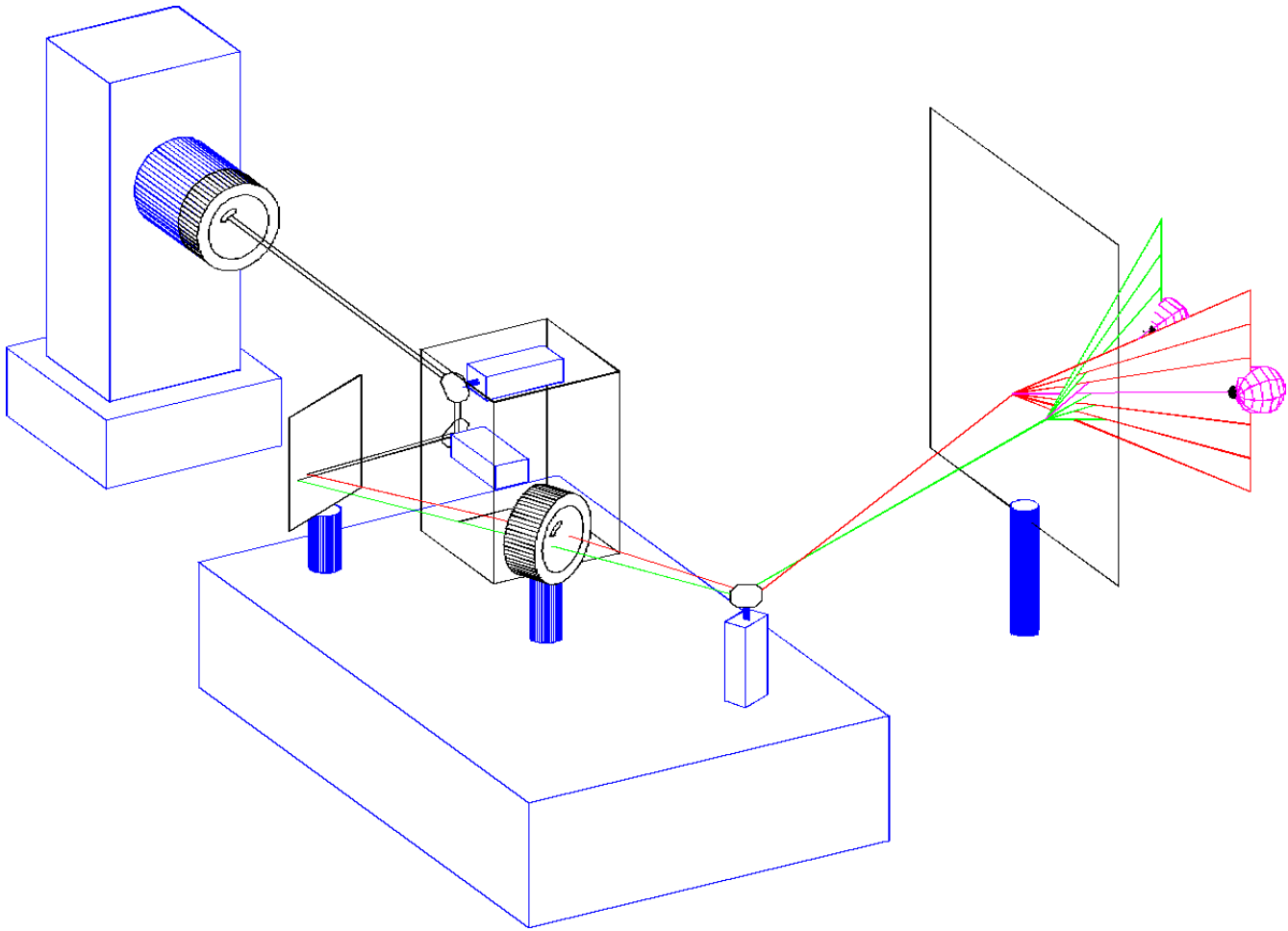

**Fig. 5** *Projection system. (b) The white light beam impinges at one side of the center of the grating, placing the voxel between the screen and the observer.*

In case b) a different positioning of the light beam gives a spectral arc at the grating and at the screen and the image is real, being represented in front of the screen at the position of intersection of both light fans. Fig. 5 allows to understand that if the filament image is precisely focused at the center of the screen it will be out of focus at both extremes of the screen. The amount of defocusing is proportional to the size of the beam when exiting from the projecting lens and to the difference in distances of its path to the center of the screen as compared to its path to one extreme. The distance from the screen to the projecting system had a variation of 15%, so that defocusing is small compared to the size of the image at the screen.

## 5. Distorsion Considerations

Addressing the points of the resulting figure laterally is free of distortion if we have precision in positioning our mirrors. The only distortion to be considered is in depth of the representation (**z coordinate**). Coordinate **z** will keep a constant relationship as long as the lens imaging process retains the value of magnification. If the distances from the spectral arcs **SE** to the screen were constant, one distortion would be present due to the difference in the impinging angle at the lateral sides of the screen. This can easily be calculated because, at each side, light impinged at an angle of 7 degrees relative to the central angle of 45 degrees, and the difference in the extension of the projection of light beams horizontally is:

$$\frac{\cos(45^0 + 7^0)}{\cos(45^0 - 7^0)} = \frac{1{,}62}{1{,}27} = 1{,}28 \tag{4}$$

that is, they are 28% larger at one side than at the other. Another distortion comes from the fact that position of the spectral arc changes at the grating, thus changing the magnification value of the projection. The amount of depth distortion will then be the sum of both effects. In our case, according to the values we figured in the next paragraph, there is a compensation of the two effects: while the angle of projection increases, increasing the spreading, there is an increase in the distance from the spectral arc to the projecting lens which reduces the magnification factor in approximately the same amount.

## 6. Experimental Details

The system was implemented by employing a 10 W, 6V halogenous lamp whose filament was oriented towards a photographic objective with a focal length of f = 80-250 mm and a numerical aperture of 4.5 . The filament was 1.45 mm thick and the first image had the magnification value of 0.5, appearing at 23 cm from the objective. **My** , the first mirror, was 7 cm from the first lens and **Mz**, the second mirror, 3 cm from **My**. The diameter of mirrors **My**, **Mz** was 2 cm and were about 15 cm away from the grating. The image was directed to a 1,200 lines/mm reflecting diffraction grating made by laser interference on photoresist material, which was developed and aluminized.

When doing the first image directed to the tilted grating, the variation allowed to the distance from the illuminated points on the grating to the projection lens was 3.5 cm. The size of the grating was 4 x 4 cm$^2$ and the beam impinged on it at $19^0$ to the grating normal. The center of the green spectral region was seen exiting at $90^0$ from its incidence direction. It entered a photographic objective PENTAX TAKUMAR whose focal length was 90 mm and numerical aperture 1:2.8 , located 6.6 cm away. A third mirror 2.5 cm high and 3.5 cm wide was close to the exit of this photographic objective reflecting the light to a transmission diffractive screen whose center was located at a distance of 1 m.

The beam exiting from the projecting lens had a vertical extension of $3 \pm 0.5$ mm and horizontal dispersion of about 32 mm. The screen was 35 cm high, 63 cm wide and tilted at an angle of $45^0$. It was made and employed as described in[11] , in AGFA holographic film 8E75 under red laser light, developed in metol-ascorbic acid developer and bleached in dichromate bleach. Diffraction efficiency is the product of the diffraction efficiency value of the holographic elements: diffraction grating and screen. We estimated this value as less than 30% , so that we have a final efficiency of less than 9%. This poor lighting of the figure is compensated by concentration of the viewing area, and again reduced due to the velocity of the flying spot. The optimum position for observation was 50 cm in front of the screen, but with tolerance it was possible to observe the scene for up to 2 m. At 2 m from the screen, there could be three simultaneous observers.

A lateral field of view of 21 cm at a 1m distance was allowed the observer, enough to see the parallax changes by moving his/her head. This field is dependent on the relative intensity of the scene related to the ambiance, and could only be achieved under conditions of almost complete darkness. But it must be indicated that, according to simple angular measurements and calculations on the diffraction spreading of wavelengths, only 160 nm of the white light spectrum reached the screen. If a new design could employ the whole visible spectrum, the field of view could be certainly doubled. The observed figure is shown in Figure 6 .

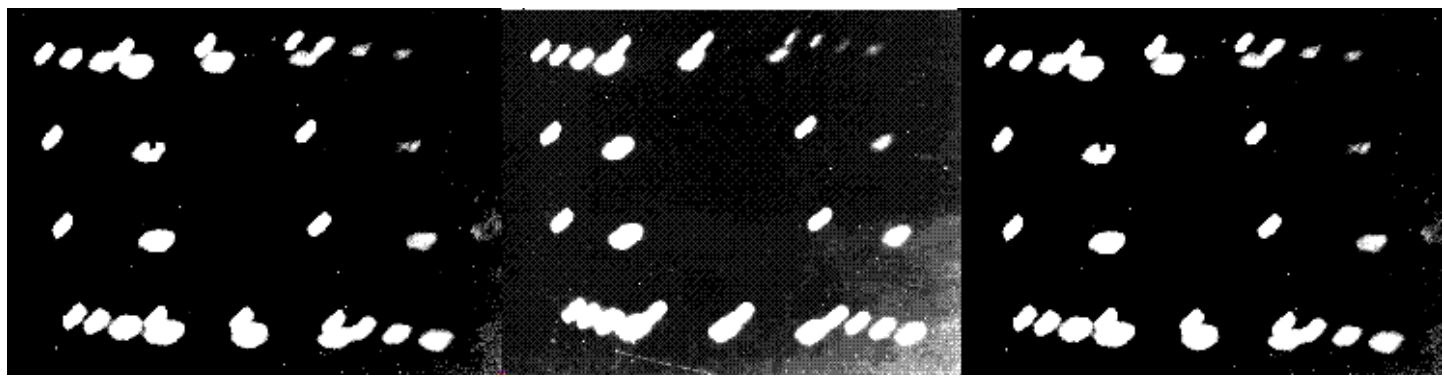

**Fig. 6** *A stereo view of the representation of a cube, in an L-R-L presentation.*

It could be seen at once, thanks to retinal persistence of the image, but some undesirable vibrations coming from the overly long supports of the mirrors destroyed the linearity of the movements. The photographs were taken not in a single long exposure shot but, due to those vibrations, by stopping the light spot at the screen at regular positions within the figure (a cube whose sides measured 24 cm) at regular intervals. The amount of defocusing for this figure had a calculated maximum of 0.2 mm .

The result is presented in a particular stereoscopic format, called L-R-L, where the classical left-right pair is incremented with another left view at its right hand side. Some individuals can watch with unassisted eyes the classical L-R stereo pair in parallel viewing, but others can only perform a cross-view sighting, which they can do in this format by selecting the R-L part of the figure.

Stepping motors were employed which had been obtained from drivers of microcomputers. Each step corresponded to $0.3^0$ because a reduction system was made by transmission belts. The object could be represented by using 23 steps in each direction. A computer program made in Pascal controlled the sequence and performed the necessary correction of steps on the x direction related to the previous position mirror $M_z$. The spot defined a resolution of less than 5 mm (the image of the filament at the screen), which appears bigger in the long exposure photographs of Fig. 6, for some spots which were brighter. When the spot was moving fast a thin line resulted. No correction was made in order to compensate the length change of the projected spectral element due to differents obliquities of the incidence, from one side of the screen to the other.

The result appears very linear to the observer. As is usual with diffraction screens acting by spectral wavelength encoding[11], a pseudoscopic image can be obtained just by changing its position symmetrically to observe the symmetric diffracting order.

**7.Conclusions**

We obtained a system which becomes a new proposal for obtaining 3D images (computer figures) in continuous horizontal parallax by electronic means, not needing massive moving components, lasers, or very large computing capability that are necessary in similar systems. They are bigger than any previously obtained similar representation and can appear in front of the screen floating in a free space that can be traversed by objects for the purpose of three-dimensional comparison with the represented image.

The result is not bright with filament lamp sources, but can be brighter with xenon lamps. LEDs , which progress rapidly in the visible range, might be suitable for this application.

We acknowledge financial contributions from the following brazilian institutions: FAPESP, UNICAMP (through FAEP) and CNPq. We are grateful to Dr. L.H. Cescato for making the diffraction grating. Leandro da Conceição de Faria and Fernando Girotto are acknowledged for helping in the measurements.